\documentclass[fleqn,12pt]{article}
\usepackage{espcrc1}

\title{A review of the hard pomeron in soft diffraction }

\author{J.R. Cudell\address{JR.Cudell@ulg.ac.be;
Institut de Physique, 
Universit\'e de Li\`ege, 
4000 Li\`ege, Belgium },
A. Lengyel\address{sasha@len.uzhgorod.ua;
Inst. of Electron Physics, Universitetska 21, UA-88000
Uzhgorod, Ukraine.},
E. Martynov\address{martynov@bitp.kiev.ua;
Bogolyubov Inst.
for Theoretical Physics, UA-03143 Kiev, Ukraine}, and
O.V. Selyugin\address{selugin@theor.jinr.ru; 
Bogoliubov Theoretical
Laboratory, JINR, 141980 Dubna, Moscow Region, Russia.}}
       
\begin{document}
\maketitle

\begin{abstract}
We review the evidence for the presence of a hard singularity 
in soft forward amplitudes, and give an estimate of its trajectory 
and couplings.
\vspace{1pc}
\end{abstract}

In 1998, Donnachie and Landshoff showed that DIS data at
small $x$ can be described by the superposition of two Regge exchanges : the
soft and the hard pomeron, and that both are well approximated by simple poles
\cite{DL98}. 
Their fit to HERA data then leads to the conclusion that the soft pomeron is a higher-twist
contribution. 
It was further suggested \cite{CDL} that the DGLAP evolution has to be
performed in a non-trivial manner : as it is of perturbative origin, 
the singularity
that it introduces at $J=1$ should be considered only if one is far away
from it, as otherwise perturbation theory breaks down. Applying
this philosophy to the evolution leads to a prediction of the $Q^2$
dependence of the hard pomeron from DGLAP evolution, and to a
successful description of $F_2$ and
$F_L$ \cite{DL01,DL03}, provided that the gluon distribution is associated
with the hard pomeron contribution.
The obtained fit can then be extended to higher $x$ \cite{DL01}.
This leads to the conclusion that the data at large $Q^2$ and $t=0$ contain
a sizable contribution from a hard pomeron with intercept
$$\alpha_h(t=0)\approx 1.4.$$

Furthermore, the same idea can be used to describe $F_2^c$ \cite{DL99a}
as well as
vector-meson photoproduction \cite{DL99b}. 
In this case, the main characteristic of the
data is a modification of the energy dependence of the cross section
as the vector-meson mass 
increases, and a flattening of the $t$ dependence at high $t$. 
Both effects can be interpreted as a signature of the contribution of
a hard pomeron, 
provided it has a rather flat trajectory:
$$\alpha_h(t)\approx 1.4+0.1 t.$$

The question remained however to understand why such a hard pomeron had
not been seen before, and what happened to it at $t=Q^2=0$. The first
remark is that it is possible to have singularities that manifest themselves 
only in photon scattering (e.g. that associated with the box diagram in $\gamma
\gamma$ scattering). Hence the first place to look is presumably the photon
total cross section. Unfortunately, the data from LEP suffer from large
theoretical uncertainties linked with the Monte-Carlo simulations 
used to unfold the data. A fit to the $\gamma\gamma$
data
only indicates that the hard pomeron $may$ be present, but with a rather small
coupling, about 10\% that of the soft pomeron \cite{DL98}. 
However, it seems rather strange that a hadronic object like a hard 
pomeron should decouple fully from total hadronic cross sections. 
One might argue that this has something to do with the point structure of
the photon coupling, but nevertheless it would be reassuring to find such
a contribution in purely hadronic cross sections.

Bounds on the hard pomeron have been known for a long time: the ratio of
its coupling
to that of $pp$ and $\bar p p$ was estimated to be less than $2\times 10^{-6}$
  \cite{C99}. Furthermore,
recent studies \cite{COMPETE} of hadronic amplitudes down to $\sqrt{s}=
5$ GeV dismiss models based on a simple-pole
pomeron, mainly because they cannot describe the
real part of the amplitude well, or equivalently because the fit
to $\rho$, the ratio of the real part to the imaginary part of 
the elastic hadronic amplitude, has an
unacceptably high $\chi^2$. 

We revisited this problem \cite{paper} first by improving the treatment
of the real part of the amplitude: 
\\
$\bullet$ We included and fitted the subtraction constant present in the
real part of the amplitude because of rising $C=+1$ contributions.
\\ $\bullet$ We used integral dispersion relations down to the correct 
threshold and, at low energies (for which the analytic asymptotic model
is not correct), we used (a smooth fit to) the data for $\sigma_{tot}$ 
to perform the dispersion integral. \\
$\bullet$ We used the exact form of the flux factor ${\cal F}=2m_p p_{lab}$
and Regge variables $\tilde s\equiv {s-u\over 2}$ proportional to
$\cos(\theta_t)$ instead of their dominant terms at large-$s$\footnote{In the $\gamma\gamma$ case, we use ${\cal F}=s$}. 

Following~\cite{COMPETE}, we fit total cross sections and $\rho$
for $pp$, $\bar p p$, $\pi^\pm p$ and $K^\pm p$, and total cross sections
for $\gamma p$ and $\gamma\gamma$ in the region $\sqrt{s}\geq$ 5 GeV.
Furthermore, as we are using simple poles, we use 
Gribov-Pomeranchuk factorisation of the
residues  
at each simple pole to predict the $\gamma\gamma$ amplitude
from the $pp$ and $\gamma p$ data~\cite{CMS}. 

If we define the hadronic $ab$ amplitude as ${\cal A}_{ab}={\Re}_{ab}+i{\Im}_{ab}$, we obtain the total cross section as
$\sigma _{tot}^{ab}\equiv \Im^{ab}/(2 m_b p_{lab})$,
with $p_{lab}$ the momentum of particle $b$ in the $a$ rest frame,
and the models that we consider are defined by the following equation:
\begin{equation}
\label{fluxfac}
{\Im}_{ab}\equiv s_1\left[ {\Im}^{R+}_{ab}\left({\tilde s\over s_1}
\right) +{\Im}^{S}_{ab}\left( {\tilde s\over s_1}\right) \mp {\Im}^-_{ab}
\left({\tilde s\over s_1}\right) \right] ,
\end{equation}
with $s_1=1$ GeV$^2$, and the $-$ sign in the last term for particles.
For the two reggeon contributions $ {\Im}^{R+}$ and ${\Im}^-_{ab}$, 
we use (non-degenerate) simple-pole expressions. For the pomeron
contribution ${\Im}^{S}_{ab}$, we allow two simple poles to contribute:
\begin{equation}
\label{poles}
\Im^S_{pb}=S_{b}\left( {\tilde s\over s_1}\right)^{\alpha _{o}}
+H_{b}\left({\tilde s\over s_1}\right)^{\alpha _{h}}
\end{equation}
For comparison, we also consider expressions corresponding to
a dipole $\Im^S_{pb}=$\break $({\tilde s/ s_1})D_{b}\log( {\tilde s}/{s_{d}})$
or a tripole
$\Im^S_{pb}=$ $(\tilde s/ s_1)T_{b}\left[\log ^{2}({\tilde s}/{s_{t}})+t'_{b}\right]$.

The improved treatment of $\rho$ leads to a better fit in all cases (a
dipole pomeron reaches a $\chi^2/dof$ of 0.94 and
a tripole pomeron  
one of 0.93, whereas they were both 0.98 in the 
standard analysis \cite{COMPETE}).
However, if we use only one simple pole for the pomeron 
($i.e.$ if we set $H_b=0$ in
(\ref{poles}), we still cannot get a fit comparable to those obtained with a 
dipole or a tripole.



However, we found that the inclusion of the second singularity in 
(\ref{poles}) has a dramatic effect: the $\chi^2$ drops from 661 to 551 for
619 points, nominally a 10 $\sigma$ effect! More surprisingly, the
new singularity has an intercept of 1.39, very close to that 
obtained in DIS. However, as was already known 
\cite{C99}, the new trajectory, which we shall call the hard pomeron,
almost decouples from $pp$ and $\bar p p$ scattering. Nevertheless, it improves
considerably the description of $\pi p$ and $K p$ amplitudes, and 
parametrisation (\ref{poles}) becomes as good as the tripole fit advocated in
\cite{COMPETE}.

The decoupling in $pp$ and $\bar p p$ scattering can easily be understood: any sizable
coupling will produce a dramatic rise with $s$, and only $pp$ and $\bar p p$ 
data reach high energy. For these data, the hard pomeron contribution needs 
to be unitarised (see however~\cite{DL04} for
a different opinion). To get a handle on the hard pomeron parameters, it is thus a good idea to fit to lower energies first. We choose to consider the region 
from 5 to 100~GeV (which includes all the $\pi p$ and $Kp$ data). We checked
that the parameters describing the hard pomeron component are stable if we 
slightly change the region of interest, by augmenting the minimum 
energy to $e.g.$ 10 GeV, or by decreasing the maximum energy to 
$e.g.$ 40 GeV.

Our best estimate for the 
hard-pomeron intercept is
\begin{equation}
\alpha_h(t=0)=1.45\pm 0.01.
\end{equation}
However, a new and unexpected hierarchy of couplings is needed. The coupling
of the hard pomeron to protons is about three times smaller than that to
pions and kaons, and is about 4\% of the coupling of the soft pomeron. 

The hard pomeron is probably not a simple pole, but it must be close to it: as we obtain the $\gamma\gamma$ cross section via Gribov-Pomeranchuk factorisation, we indeed test the analytic nature of the singularity \cite{CMS}. 
The LEP data are compatible with our results, 
and we prefer a lower value, such as that obtained using PHOJET. Note that
the fit of \cite{DL04}, which has a much smaller hard pomeron coupling
because the simple-pole structure is used up to the Tevatron, leads to
a larger $\gamma\gamma$ cross section, compatible this time with the
data unfolded with PYTHIA.
Hence our value of the coupling of the hard pomeron to protons must
be an upper limit: bigger values would lead to too small 
a $\gamma\gamma$ cross section, whereas that of \cite{DL04} would be
a lower limit.

To extend our fit to higher energies, one must unitarise the hard
pomeron contribution, as it violates the black-disk limit around $\sqrt{s}=400$ GeV. The way to do this is far from clear, especially as there can be some mixing with the other trajectories.
We have shown in~\cite{paper} that it is possible to find a unitarisation scheme
which produces a good description of the data for all energies. 
The contribution if the hard pomeron is then always smaller than 25\% of the total
cross section. 

In conclusion, several independent analyses point to the fact that a hard
pomeron may exist, both in DIS, photoproduction and soft cross sections.
This object may be similar to a simple pole for $\sqrt{s}\leq 100$ GeV,
and its coupling to protons is small for soft cross sections.
Its contribution is important
at large $s$, large $Q^2$ or large $t$. 
The surprising hierarchy of couplings, as well as their
smallness, indicate that our results need confirmation. 
It may be worth noting here that we obtain similar results
if we exclude the $\rho$ data from the analysis. 
Finally, we must insist on the fact that the combination of two simple
poles (a soft and a hard pomeron) is only one of several possibilities.
For soft data and DIS data, other models exist \cite{S} 
which are also compatible
with unitarity, and which produce equally good fits.
A study of elastic scattering \cite{coming} may help distinguish
the two-simple-pole model from the tripole and the dipole, 
as it predicts a contribution with a fast rise but a small slope, which
is not natural in the competing models.

\end{document}